\documentclass[aps, twocolumn, letterpaper, prl]{revtex4}

\usepackage{amsmath}
\usepackage{amssymb}
\usepackage{graphicx}
\usepackage{xspace}
\usepackage{upgreek}
\usepackage{accents}

\newcommand{\eg}{{e.g.,\/}\xspace}
\newcommand{\ie}{{i.e.,\/}\xspace}

\newcommand{\eq}[1]{(\ref{#1})}
\newcommand{\Eq}[1]{Eq.~(\ref{#1})}
\newcommand{\Eqs}[1]{Eqs.~(\ref{#1})} 
 
\newcommand{\Ref}[1]{Ref.~\cite{#1}}
\newcommand{\Refs}[1]{Refs.~\cite{#1}}

\newcommand{\mc}[1]{\mathcal{#1}}
\newcommand{\mcc}[1]{\mathfrak{#1}}

\newcommand{\favr}[1]{\langle #1 \rangle}
\renewcommand{\vec}[1]{{\boldsymbol{\rm #1}}}
\newcommand{\oper}[1]{\hat{\vec{#1}}}
\newcommand{\pd}{\partial}
\newcommand{\avp}{{f}}

\newcommand{\kpt}[1]{{\kern #1 pt}}

\newcommand{\msection}[1]{\textit{#1.}\ ---\ }  

\begin{document}
\title{Ponderomotive forces and wave dispersion: two sides of the same coin}
\author{I.~Y. Dodin}
\affiliation{Department of Astrophysical Sciences, Princeton University, Princeton, New Jersey 08544, USA}
\date{\today}

\begin{abstract}
Presented here is a general view on adiabatic and resonant wave-particle interactions leading to a uniform description of nonlinear ponderomotive effects in very different environments, from low-temperature plasmas to relativistic plasmas or even atoms in laser light. Treating the wave-particle interaction as a classical mode-coupling problem, this theory shows the inherent connection between the ponderomotive forces and the properties of waves causing those forces. The adiabatic Lagrangians are derived for single particles and nonlinear waves, possibly carrying trapped particles, and yield both the dynamic equations and the nonlinear dispersion relations in the general case.
\end{abstract}

\maketitle

\msection{1. Introduction} A particle traveling in a rapidly oscillating wave field attains an average interaction energy $U$ acting as an effective ``ponderomotive'' potential. In low-temperature plasmas, $U$ easily can be made comparable or larger than the particle kinetic energy, thus producing \textit{barriers}, which are generally asymmetric, irreversible, and exhibit quantumlike properties \cite{my:invited}. It was found recently that certain features of these barriers are akin to wave-particle interactions in also very different environments, from $\alpha$ channeling in fusion plasmas to laser traps for atomic cooling \cite{my:invited, my:dipole, my:beatf, my:nlinphi} and even relativistic laboratory and cosmological plasmas \cite{my:mneg, my:ebunch, my:mquanta}. Elaborating on these analogies could help understand the key underlying physics and also facilitate sharing techniques of wave and particle manipulation between different regimes. Thus, a unifying theory of ponderomotive interactions is needed.

In \Refs{my:kchi, my:mneg, my:nlinphi, my:bgk, my:meffham}, a suitable analytical framework was developed by approaching wave-particle interactions as a variety of the classical mode-coupling problem. In this tutorial we highlight most general aspects of this framework, leaving out literature review and discussions on effects specific to particular physical systems. For those, one is referred to our earlier publications, and more explanatory papers will also follow.

\msection{2. Single particle in a given field} Let us start out with considering a particle in a given wave field, assuming that collisions and dissipation can be neglected. (If necessary, those can be included later as perturbations \cite{my:mefffric}.) Then the particle trajectory is found from the least-action principle: among trajectories $\vec{x}(t)$ starting at $\vec{x}_1$ at time $t_1$ and ending at $\vec{x}_2$ at time $t_2$, realized is the one on which the action $\int^{t_2}_{t_1} L\,dt$ is minimal. Here $L$ is the particle Lagrangian, which we represent as $L = \favr{L} + L_\sim$, with $\favr{L}$ being the local time-average over the particle oscillations, and $L_\sim$ being the quiver part. On time scales large compared to the oscillation time scales, it is only $\favr{L}$ that contributes to the integral; hence, $\Lambda \equiv \favr{L}$ acts as the slow-motion Lagrangian. 

To make use of this theorem, one needs to decompose the particle coordinate and velocity $(\vec{x}, \vec{v})$ into the slow variables $(\vec{X}, \vec{V})$ and the quiver variables $(\tilde{\vec{x}}, \tilde{\vec{v}})$, express the latter in terms of $(\vec{X}, \vec{V}, t)$ \cite{foot:accuracy}, substitute the result into $L(\vec{x}, \vec{v}, t)$, and perform the time-averaging to obtain $\Lambda(\vec{X}, \vec{V}, t)$, the time dependence retained being slow. Then, the motion equations are yielded in the form of Euler-Lagrange equations as usual; see, \eg \Ref{my:meff}.

In general, the particle itself is an oscillator, \ie has degrees of freedom additional to that of the translational ``oscillation center'' (OC) motion $(\bar{\vec{x}}, \bar{\vec{v}})$. (Examples here are gyromotion, intramolecular vibrations, etc.) Hence, it can be assigned one or more internal frequencies $\Omega = \dot{\varphi}$, with $\varphi$ being the corresponding canonical phase. If these oscillations are fast enough, the explicit dependence on $\varphi$ in the Lagrangian vanishes at time-averaging, so $\Lambda$ can be written as a function of $(\bar{\vec{x}}, \bar{\vec{v}}, \Omega, t)$; cf. \Ref{my:nlinphi}. Then, the canonical action $J \equiv \pd_{\Omega} \Lambda$ is an adiabatic invariant, so one can reformulate the variational principle such that the internal degrees of freedom are omitted. Namely \cite{my:mneg}, among trajectories $\bar{\vec{x}}(t)$ starting at $\bar{\vec{x}}_1$ at time $t_1$ and ending at $\bar{\vec{x}}_2$ at time $t_2$, \textit{with arbitrary} $\varphi_1$ \textit{and} $\varphi_2$, realized is the one on which the action $\int^{t_2}_{t_1} \mc{L}\,dt$ is minimal. Here, the reduced Lagrangian $\mc{L}$, or the Routhian \cite{book:landau1}, reads as
\begin{gather}
\mc{L} = \Lambda - J \Omega, \quad  \Omega = \Omega(J, \bar{\vec{x}}, \bar{\vec{v}}, t),
\end{gather}
so it is a function of $(\bar{\vec{x}}, \bar{\vec{v}}, t)$ and depends on $J$ parametrically. Then, the OC Hamiltonian equals
\begin{gather}
\mc{H} = \vec{P}\cdot\bar{\vec{v}} - \mc{L},
\end{gather}
where $\vec{P} = \pd_{\bar{\vec{v}}}\mc{L}$ is the canonical momentum \cite{my:mneg}. In particular, $d_t \mc{H} = \pd_t\mc{H}$, so, in a stationary wave, $\mc{H}$ is an adiabatic invariant too \cite{my:beatf}.

Notice that $\mc{L}$, $\vec{P}$, and $\mc{H}$ can be of non-Newtonian form already in the nonrelativistic limit, thus giving rise to a plethora of unusual effects, \eg the negative-mass effect \cite{my:mneg, tex:mytrieste, my:meffham}. Still, the same formalism is also applicable in fully relativistic calculations \cite{my:mneg, my:meff, my:gev}.

\msection{3. Invariants at resonant interactions} Suppose now that the particle interacts resonantly with a stationary wave, thereby contributing a periodic perturbation $\mc{H}_\sim$ to $\mc{H}$. Then $J$ and $\mc{H}$ are no longer conserved separately; yet, another general invariant can be found. To see this, let us extend the OC phase space by formally adding another degree of freedom $(\theta, I)$ corresponding to the oscillations at the wave frequency $\omega$. Hence the explicit time dependence in $\mc{H}_\sim$ is replaced with the dependence on the phase $\theta$, and for the action variable one obtains $I = - \mc{H}/\omega$ \cite{my:beatf}. Now our system formally contains two oscillators, $(\Omega, J)$ and $(\omega, I)$, so the generalized Manley-Rowe theorem applies \cite{my:manley}. Specifically, suppose the resonance condition in the form, say, $n_\omega \omega \approx n_\Omega \Omega$, where $n_\omega$ and $n_\Omega$ are some integers. Then, although $J$ and $I$ are no longer conserved separately, they are yet constrained to the so-called diffusion path \cite{my:beatf, ref:tennyson82}, $dI/n_\omega = - dJ/n_\Omega$, yielding
\begin{gather}\label{eq:dp}
\mc{H} - (n_\omega/n_\Omega)\, \omega J = \text{const}.
\end{gather}

Equation \eq{eq:dp} generalizes the integral found in \Refs{my:cdlarge, my:ratchet} for asymmetric barriers in low-temperature plasmas and is also key to the cooling mechanism proposed in \Refs{my:invited, my:dipole}. The latter is similar to the celebrated Sisyphus cooling \cite{ref:cohen-tannoudji98}, showing inherent connections of the present theory with the physics of laser-atom interactions \cite{my:dipole}. Yet, remarkably, the same conservation law \cite{my:beatf} also facilitates $\alpha$ channeling in fusion plasmas \cite{ref:fisch92} and thus suggests that akin techniques can be used to manipulate particles in very different environments.

\msection{4. Ponderomotive forces from dispersion} The angle-action representation for the wave variables turns out to be convenient also for calculating the ponderomotive Hamiltonian $\mc{H}$ itself, even when the particle dynamics is considered in a prescribed classical field. For example, consider the wave as a mode with a prescribed spatial structure and suppose, for simplicity, that it is coupled adiabatically to a single particle. Then, $J$ and $I$ are conserved (the latter henceforth defined as the wave true action, unlike in Sec.~3), so what is affected by the interaction are the \textit{eigenfrequencies} of the wave-particle system. In \Ref{my:kchi}, we showed that the shifts of these frequencies satisfy
\begin{gather}\label{eq:domega}
\delta \Omega = (\pd_J U)_{\bar{\vec{v}}},
\quad
\delta \omega = (\pd_I U)_{\bar{\vec{v}}},
\end{gather}
where $U \equiv \mc{L} - \mc{L}_0$ is the interaction Lagrangian, and $\mc{L}_0$ is the particle OC Lagrangian in the absence of the wave. (Particularly notice that the derivatives are taken at fixed~$\bar{\vec{v}}$.) This means that knowing the wave microscopic frequency shift $\delta \omega$ per particle, one can, in principle, find $U$. In other words, knowing the wave dispersion is sufficient to derive the ponderomotive force.

In the simplest case when the interaction is linear, $\delta \omega$ is independent of $I$, so one obtains $U \approx \Phi$, where $\Phi = I\,\delta \omega$ is called the dipole ponderomotive potential. Since $I$ per unit volume equals $\mc{E}/\omega$, where $\mc{E}$ is the wave energy density, one thereby obtains that, for species $s$, the ponderomotive potential $\Phi_s$ is proportional to the derivative of $\omega$ with respect to the particle density~$n_s$:
\begin{gather}\label{eq:phid}
\Phi_s = \frac{\mc{E}}{\omega}\,\frac{\pd \omega}{\pd n_s}.
\end{gather}

Equation \eq{eq:phid} proves useful, \eg for calculating the effect of ionization and recombination on homogeneous waves in cold plasma \cite{my:iorec}. Yet, it can also be used to obtain $\Phi$ in terms of the local field variables rather than those of an actual mode. Namely, substituting $\mc{E}$ and the general dispersion relation for $\omega$, one obtains \cite{my:kchi}
\begin{gather}\label{eq:phi}
\Phi = - {\scriptstyle \frac{1}{4}}\,\tilde{\vec{E}}^*\cdot \oper{\alpha} \cdot \tilde{\vec{E}},
\end{gather}
where $\tilde{\vec{E}}$ is the electric field complex amplitude, and $\oper{\alpha}$ is the particle linear polarizability. (Alternatively, \Eq{eq:phi} is obtained by assigning to the particle a dipole moment $\oper{\alpha}\cdot\tilde{\vec{E}}$, which characterizes the charge displacement from the OC trajectory \cite{my:nlinphi}.) This also yields that
\begin{gather}\label{eq:hdip}
\mc{H} \approx \mc{H}_0 + \Phi,
\end{gather}
where $\mc{H}_0$ is the OC Hamiltonian in the absence of the wave. In other words, the second-order (in $\tilde{E}$) interaction Hamiltonian is determined by the \textit{linear} polarizability, the statement being known as the $K$-$\chi$ theorem. In this approximation, the wave-particle energy is also separated automatically into the wave energy and the OC energy, with zero coupling between the two \cite{my:kchi}.

If a particle is an oscillator, $\oper{\alpha}$ exhibits a singularity near the resonance. Namely, at $\omega \approx \Omega$, one obtains $\Phi \propto |\tilde{E}|^2/(\omega - \Omega)$, always with a positive coefficient \cite{my:nlinphi}. A nonlinear oscillator will yield a hysteretic $U$, as one finds similarly by integrating \Eq{eq:domega} for a model $L$ \cite{my:nlinphi}. Finally, if multiple internal modes are present, coupling at beat resonances yields $U$ matching the effective \textit{quantum} potential seen by a two-level atom in a laser field. For example, in the simplest case of large enough detuning $\Delta = \omega - (\Omega_1 - \Omega_2)$, the latter reads as
\begin{gather}
U = \frac{\hbar \Omega_{\rm R}^2}{4\Delta}\,(n_2-n_1),
\end{gather}
where $\Omega_{\rm R}$ is the Rabi frequency, and $n_i$ are the corresponding occupation numbers \cite{my:nlinphi}.

Remarkably, these results are not limited to specific physical systems and ensure that ponderomotive forces in the absence of dissipation are of manifestly Lagrangian or Hamiltonian form. Hence, such forces conserve phase space automatically, unlike when they are derived \textit{ad hoc} from time-averaging of the motion equations.

\msection{5. Dispersion from ponderomotive forces} One can also invert the argument and derive $\omega$ from $U$ instead. This can be done by summing contributions of individual particles in the right-hand side of \Eq{eq:domega}, so a differential equation for $\omega$ is yielded. Alternatively, the wave dispersion is obtained in a nondifferential form as follows. Along the lines of Sec.~2, the wave Lagrangian (Routhian) density is derived in the geometrical-optics approximation \cite{my:bgk, tex:mytobe}, reading~as
\begin{gather}\label{eq:lagrw}
\mcc{L} = \favr{\mcc{L}_{\rm em}} - \sum_s n_s \favr{\mc{H}_s}_{\avp}.
\end{gather}
Here $\favr{\mcc{L}_{\rm em}} = \favr{E^2 - B^2}/(8\pi)$ is the time-averaged Lagrangian density of the electromagnetic field, with electric and magnetic fields $\vec{E}$ and $\vec{B}$ including both wave and quasistatic fields, if any; summation is taken over distinct species $s$, and $\favr{\mc{H}_s}_{\avp}$ are the corresponding OC Hamiltonians averaged over the local distributions~$f_s$. 

Understanding the wave properties hereby becomes straightforward. Namely, to obtain $\mcc{L}$, one needs to calculate \textit{only} the single-particle Hamiltonians $\mc{H}_s$. The latter is done along the lines of Sec.~2, thus rendering the Maxwell-Vlasov system unnecessary in many cases of interest, unlike in other existing theories \cite{my:bgk}. In the simplest case, one can use the ponderomotive Hamiltonian in the dipole limit [\Eqs{eq:phi} and \eq{eq:hdip}]; then,
\begin{gather}
\mcc{L} = \mcc{L}_0 + \frac{1}{16\pi}\,\left(\tilde{\vec{E}}^* \cdot \oper{\epsilon} \cdot \tilde{\vec{E}} - |\tilde{\vec{B}}|^2\right),
\end{gather}
with $\mcc{L}_0$ independent of the field variables. (Here we used that the linear dielectric tensor $\oper{\epsilon}$ equals $1 + 4\pi \sum_s n_s \favr{\oper{\alpha}_s}_{\avp}$.) Hence, nondissipative linear waves are recovered immediately \cite{my:bgk}. Yet, \Eq{eq:lagrw} also describes nonlinear waves, nonperturbatively in the field amplitude $a$. The general dispersion relation is obtained by varying $\mcc{L}$ with respect to $a(\vec{x}, t)$; specifically,
\begin{gather}\label{eq:ndr}
\pd_a \mcc{L} = 0.
\end{gather}

The envelope equation, or the action conservation theorem \cite{foot:act} can be derived as well, by varying $\mcc{L}$ with respect to the wave phase~$\xi(\vec{x}, t)$. In the simplest case, Whitham's equation is recovered \cite{ref:whitham65}:
\begin{gather}\label{eq:act}
\pd_t (\pd_\omega \mcc{L}) - \nabla \cdot (\pd_{\vec{k}} \mcc{L}) = 0.
\end{gather}
However, we found that it can also be altered, namely, in the presence of particles trapped by the wave \cite{tex:mytobe}. A peculiar feature of such particles is that their OC locations $\bar{\vec{x}}$ are determined by the motion of the wave nodes and thus the trapped-particle density is connected with $\xi$. Also, the wave Lagrangian density $\mcc{L}$ attains a term that, albeit being a function of $\omega$ and $\vec{k}$, is independent of $a$. This special property of trapped-particle waves results in effects that may not be captured correctly by other existing theories \cite{tex:mytobe}. More explanatory papers will follow that will describe our specific findings in this area.

\msection{6. Summary} Presented here is a general view on adiabatic and resonant wave-particle interactions leading to a uniform description of nonlinear ponderomotive effects in very different environments, from low-temperature plasmas to relativistic plasmas or even atoms in laser light. Treating the wave-particle interaction as a classical mode-coupling problem, this theory shows the inherent connection between the ponderomotive forces and the properties of waves causing these forces. Both are attributed to the same effect, namely, the eigenfrequency shifts in the wave-particle system. In particular, it is made clear how \textit{nonlinear} ponderomotive energy in the dipole limit is responsible for the \textit{linear} dispersion, and vice versa. Yet, the same formulation resolves also essentially nonlinear effects, \eg higher-order ponderomotive forces and effects of trapped particles on the wave properties in plasma.

\msection{Acknowledgments} The work was supported through the NNSA SSAA Program through DOE Research Grant No. DE274-FG52-08NA28553.


\begin{thebibliography}{10}
\bibitem{foot:accuracy}
The lowest-order average forces will be retained this way even if
  $(\tilde{\vec{x}}, \tilde{\vec{v}})$ are calculated under the approximation
  that the wave field is locally homogeneous and stationary.
\end{thebibliography}

\end{document}